\documentclass[aps,pre,reprint,floatfix,superscriptaddress,amsfonts,amsmath,amssymb,citeautoscript]{revtex4-2}
\usepackage[pdftex]{graphicx}

\begin{document}

\title{Effective medium theory for the electrical conductance of random resistor networks which mimic crack-template-based transparent conductive films}

\author{Yuri Yu. Tarasevich}
\email[Corresponding author: ]{ytarasevich@id.uff.br}
\affiliation{Instituto de F\'{\i}sica, Universidade Federal Fluminense, Niter\'{o}i, RJ, Brasil}

\author{Irina V. Vodolazskaya}
\email{vodolazskaya\_agu@mail.ru}
\affiliation{Laboratory of Mathematical Modeling, Astrakhan State University, Astrakhan, Russia}

\author{Andrei V. Eserkepov}
\email{dantealigjery49@gmail.com}
\affiliation{Laboratory of Mathematical Modeling, Astrakhan State University, Astrakhan, Russia}

\author{F\'{a}bio D. A. Aar\~{a}o Reis}
\email{fabioreis@id.uff.br}
\affiliation{Instituto de F\'{\i}sica, Universidade Federal Fluminense, Niter\'{o}i, RJ, Brasil}

\date{\today}

\begin{abstract}
We studied random resistor networks produced with regular structure and random distribution of edge conductances. These networks are intended to mimic crack-template-based transparent conductive films as well some random networks produced using nano-imprinting technology. Applying an effective medium theory, we found out that the  electrical conductance of such networks  is  $\approx 0.5852 \sqrt{n_E}$, where $n_E$ is the number density of conductive edges. This dependence is in agreement with numerical calculations in Voronoi networks, although the effective conductances are approximately 15\% larger. 
\end{abstract}

\maketitle

\section{Introduction\label{sec:intro}}

Transparent conductive films (TCFs) also denoted as transparent conductive electrodes (TCEs) are widely used for creating displays and touch screens~\cite{Gao2016}, transparent heaters~\cite{Gupta2016,Papanastasiou2020}, solar cells~\cite{Rao2014,Muzzillo2017,Haverkort2018,Zhang2020,Muzzillo2020}, electromagnetic interference (EMI) shields~\cite{Han2016,Walia2020,Voronin2023}, thermochromic devices~\cite{Cui2020}, etc. Widespread devices based on oxides, e.g., zinc oxide (ZnO), indium tin oxide (ITO), etc., have a number of significant disadvantages~\cite{Gao2016}. For example, there are high material costs of ITO-based TCFs, indium deficiency, damage to organic substrates during sputtering, and fragility; an important disadvantage for their application in solar cells is the strong absorption of ITO in the UV and blue spectral ranges~\cite{Groep2012}; brittleness is a serious obstacle to the use of such transparent electrodes for flexible and stretchable electronics~\cite{Langley2013,Ko2013,Sannicolo2016,Guo2015,Wei2019}. The listed disadvantages lead to the search for new approaches, for example, the development of printing technologies~\cite{Li2018,Mou2018,Khinda2019} and the creation of transparent electrodes based on templates, including natural templates and crack templates~\cite{Han2014,Han2014a,Gao2017,Cheuk2016}.

Despite significant advances in the technology for the production of a variety of TCFs, there remain many technological and theoretical problems that are pending. In the case of TCFs based on carbon nanotubes (CNTs) and metallic nanowires (NWs), the concentration of conducting particles must be large enough to ensure the appearance of a conducting network connecting the opposite boundaries of the film, i.e., the system must be above the percolation threshold. At the same time, the concentration should be low enough to ensure high transparency of the film. These  demands are conflicting. 

In the case of nanowire-based TCFs~\cite{Khanarian2013}, overheated areas (hot spots) may appear  leading to degradation of the conductive films, which is a serious problem~\cite{Khaligh2013,Khaligh2017,Sannicolo2018,Fantanas2018,Patil2020,Charvin2021}. Heat uniformity is a key requirement in the case of transparent heaters~\cite{Gupta2017,Papanastasiou2020}. Moreover, in the case of nanowire-based TCFs, there are dead ends of the nanowires which do not contribute to the electrical conductance, while decreases the transparency of TCFs~\cite{Kumar2016}. The electrical resistance of junctions between nanowires contribute significantly to the resistance of the such TCFs~\cite{Bellew2015,Kim2019,Benda2019}. To ensure high-quality contacts between nanowires and reduce the junction resistance, various processing methods are used (see, e.g, Ref.~\onlinecite{Ding2020} with a review of technologies for welding silver nanowires). Using these technologies, depending on the initial resistance, the sheet resistance can be reduced by several orders of magnitude to tens of ohms. So-called seamless or junction-free networks are more or less free of above issues as hot spots, dead ends and junction resistance. 

Uniform illumination is crucial for ensuring the imaging quality when TCFs are used for EMI shielding in the optical imaging domain. As compared to regular periodic networks (square, honeycomb), the stray light energy from high-order diffractions by the random network is significantly less~\cite{Li2023,Li2023a}, which indicates the good optical performance of such random networks~\cite{Liu2016,Han2016,Shen2018,Jiang2019,Song2023,Li2023}. Moreover, random metal networks produce neither moiré nor starburst patterns unlike networks with periodical structure~\cite{Shin2016,Li2017,Jung2019}. 

Seamless random metallic networks produced using crack templates seem to be a very promising kind of TCFs since inhomogeneity, dead ends and hot spots are less likely, while technologies are well elaborated~\cite{Chen2024,Voronin2023}. To mimic crack templates, Voronoi tessellation (also known as Thiessen polygons) is used~\cite{Zeng2020,Kim2022,Tarasevich2023,Esteki2023,Tarasevich2023a}. Moreover, Thiessen-polygon metal meshes can be directly  fabricated through nano-imprinting technology~\cite{Song2023}. 

\citet{Kumar2016} proposed a formula describing the sheet resistance of random resistor network (RRN) on main physical parameters, viz,
\begin{equation}\label{eq:Kumar}
  R_\Box = \frac{\pi\rho}{2 A \sqrt{n_\text{E}}},
\end{equation}
where $\rho$ is the resistivity of the material, $A$ is the crossection of the wire, and $n_\text{E}$ is the number of wire segments per unit area. Although the authors considered their method as purely geometrical, in fact, this is a kind of mean-field approach (MFA). In the case of RRNs, the MFA deals with only one conductor placed in the mean electric field that all other conductors produce, instead of considering of all the conductors in a system. The approach is based on the fact that in a dense, homogeneous and isotropic two-dimensional RRN, the electric potential is expected to change linearly between two electrodes applied to opposite boundaries of such an RRN. \citet{Kumar2016} supposed that when a potential difference is applied to the opposite border of a sample, the number of conductive wires intersecting an equipotential line is supposed to be $\sqrt{n_E}$ per unit length. Recently, \citet{Tarasevich2023} demonstrated that this assumption overestimates the intersection number by about 1.5 times. More exact formula based on a MFA 
\begin{equation}\label{eq:Rsheet}
R_\Box = \frac{2 \rho}{ n_\text{E}  \langle l \rangle A}
\end{equation}
additionally includes the mean length of conductive wires, $ \langle l \rangle$~\cite{Tarasevich2023}. Although \eqref{eq:Rsheet} agrees better with the results of direct calculations of electrical conductance than~\eqref{eq:Kumar}, both formulas overestimate the electrical conductivity. 

However, the effective medium theory (EMT)~\cite{Bruggeman1935} is often applied to predict physical properties, e.g., electrical conductance, of random systems including RRNs~\cite{Kirkpatrick1973,Clerc1990,Luck1991,OCallaghan2016}.
The goal of the present work is an investigation of the electrical properties of artificial computer-generated networks that are intended to mimic the properties of real-world crack-template-based TCFs, viz, we intend to obtain a dependence of the electrical conductance of networks under consideration on main physical parameters.

The rest of the paper is constructed as follows. Section~\ref{sec:methods} describes technical details of simulation. Section~\ref{sec:results} presents the analytical approach, together with our main findings. Section~\ref{sec:concl} summarizes the main results. Some mathematical details are presented in Appendix~\ref{sec:appendixiter}.

\section{Methods\label{sec:methods}}

Results of image processing of published photos of crack-template-based TCFs evidenced that, in crack patterns, the typical value of the node valence is 3~\cite{Tarasevich2023}. A small amount of nodes owning valence 1 corresponds to dead ends, their incident edges do not contribute to the electrical conductivity. Besides, boundaries of photos produce fictional nodes with valence 1, these fictional nodes correspond none real nodes. A small amount of nodes owning valence 2, in fact, correspond to bends on wires. A small amount of nodes owning valence greater than 3 should be treated as an artifact of image processing of photos with  modest resolution, when two or more nodes are treated as only one node,  since simplest mechanical  arguments suggest that X-shaped cracks are highly unlikely. Thus, networks with valence 3 can be used to mimic crack-template-based TCFs.

A Voronoi tessellation is a partition of a space into regions close  with respect to the Euclidean distance to each of a given set of objects (see, e.g.,~\cite{Okabe2000}). In our study, we deal with random plane Voronoi diagrams. In this particular case, points (seeds) are randomly distributed within a bounded domain on a plane. In a random plane Voronoi diagram,  the degree of each vertex is~3, while the average number of vertices in a cell is 6~\cite{Meijering1953}. Thus, from the point of view of the graph theory, a random plane Voronoi tessellation generates a 3-regular planar graph. Since crack patterns demonstrate the similar property, random plane Voronoi diagram seems to be a reasonable and useful mathematical model of crack patterns~\cite{Zeng2020,Kim2022,Tarasevich2023,Esteki2023,Tarasevich2023a}. However, even simpler 3-regular network seems to be appropriate for a preliminarily evaluation of the electrical conductivity of crack-template-based TCFs, i.e., a hexagonal network in which each edge posses a random electrical resistance drawn from an appropriate distribution. Since the electrical resistance of a wire is proportional to its length, the `appropriate distribution' in this context means `the edge length distribution of random plane Voronoi diagram'.

Thus, in our study, we use a hexagonal network, the edges of which have random electrical resistances. The probability density function (PDF) of these random resistances corresponds to the PDF of edge lengths in a random plane Voronoi network. 
We used a rectangular domain $L_x \times L_y$, where $L_y = 32$, while $L_x =  L_y \sqrt{3}/2$. When the side of the regular hexagon is $a = L_y/(2N)$, the number density of hexagons, i.e., the seed concentration is $(2\sqrt{3} a^2)^{-1}$. 
For a random plane Voronoi tessellation,  the PDF of the edge lengths is known in quadratures~\cite{Muche1996,Brakke2005}, the PDF for the unit seed density was calculated~\cite{Brakke2005}. For an arbitrary seed density, the PDF can be rescaled as follows. Let the seed density be equal to $q^2$, then the new PDF can be obtained from the initial one by replacing  $l \to l/q$ and $f(l) \to qf(l)$.

\begin{figure}[!htb]
  \centering
  \includegraphics[width=\columnwidth]{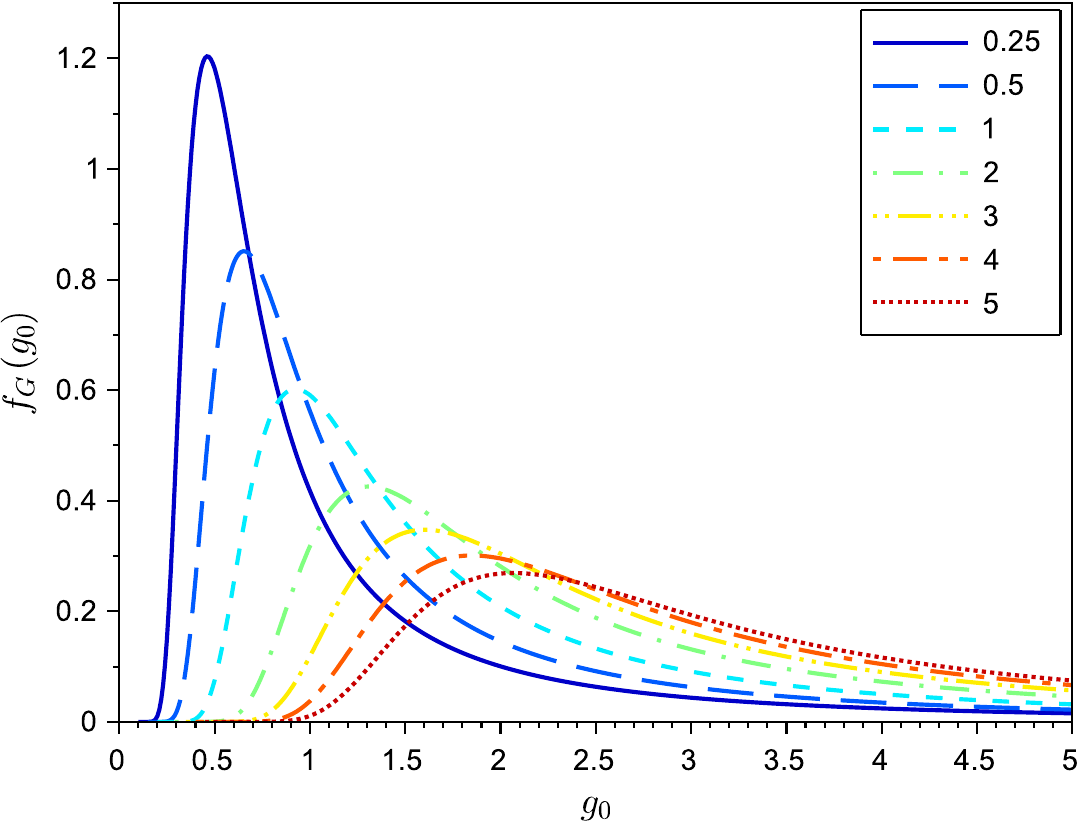}%
  \caption{PDF of edge conductivity $g$ of RRNs produced using random plane Voronoi tessellations~\eqref{eq:appfG} for different values of the seed density. For unit seed density, the values of length PDF were taken from~\cite{Brakke2005}, while, for other seed densities,  they were obtained using a scaling. $g_1 =1$.\label{fig:fvsl}}
\end{figure}

When the edge length $l$ has a PDF $f_l(x)$, the PDF of the electrical conductance  $f_G(x)$ can be found, taking into account that the electrical conductance $g_0$ depends on the length of the conductor $l$, i.e., $g_0 = \varphi(l)$, namely, the electrical conductance is inversely proportional to the length of the conductor (edge)
\begin{equation}\label{eq:appphi}
  \varphi(l) = \frac{g_1}{l},
\end{equation}
where $g_1$ is the electrical conductance per unit length. Then (see, e.g.,~\cite{Wentzel1987en}),  $ f_G(x) = \left|\psi'(x)\right| f_l\left( \psi(x) \right).$
Here, $\psi(x)$ is the inverse function of $\varphi(x)$: $\psi\left(\varphi(x)\right) = x$.
\begin{equation}\label{eq:appl}
  l = \frac{g_1}{g_0}.
\end{equation}
\begin{equation}\label{eq:apppsiprime}
  \left|\psi'(g_0)\right| = \left|\left(\frac{g_1 }{g_0}\right)'\right| = \frac{g_1}{g_0^2}.
\end{equation}
\begin{equation}\label{eq:appfG}
  f_G(g_0) = \frac{g_1}{g_0^2} f_l\left( \frac{g_1}{g_0} \right).
\end{equation}
Figure~\ref{fig:fvsl} demonstrates that the PDF~\eqref{eq:appfG} is wide, hence, a significant inaccuracy of EMT prediction is expected~\cite{Marchant1979}.

To calculate the electrical conductance, we attached a pair of superconducting buses to the two opposite boundaries of the domain in such a way that the potential difference was applied either along axis $x$ or along axis $y$. Applying Ohm’s law to each resistor (hexagon edge) and Kirchhoff's point rule to each junction (vertex of the honeycomb lattice), a system of linear equations was obtained. This system was solved numerically. Since $L_x \ne L_y$, the resistances along $x$ and $y$ axes are different. The effective conductivity, $G$, can be calculated as follows.
\begin{equation}\label{eqGeff}
  G = \frac{L_x}{R_x L_y}, \quad G = \frac{L_y}{R_y L_x}.
\end{equation} 
For each value of the seed concentration, the effective conductivity was averaged over 100 independent runs and both directions. The standard deviations of the mean are of the order of the marker size.

\section{Results\label{sec:results}}

\subsection{Effective medium theory\label{subsec:EMT}}
The main ideas of applying the EMT to networks with a regular structure and various conductances of edges are presented in the works~\cite{Kirkpatrick1971,Kirkpatrick1973,Joy1978,Joy1979,Clerc1990}. Alternatively, a more formal and general consideration based on Foster's theorem~\cite{Foster1961} is possible~\cite{Marchant1979}. Kirkpatrick~\cite{Kirkpatrick1971,Kirkpatrick1973} considered a random electrical network constructed from a hypercubic lattice in any dimension $d > 1$ (square for $d = 2$, cubic for $d = 3$, etc.). Lattice edges were treated as circuit elements with (possibly complex) independent random conductances drawn from a general probability distribution. The EMT offers a way to replace a random network with a uniform one, where all edges have the same conductance~\cite{Kirkpatrick1973}. When a voltage is applied along one axis of a RRN, the distribution of potentials in this network may be regarded as a superposition of an homogeneous `external field' and a fluctuating `local field', whose average value over any sufficiently large region has to be zero. When the values of the edge conductances are drawn from the probability distribution $f(g)$ (which can be either continuous or discrete), the requirement that the mean value of the `local field' be equal to zero gives the condition that determines the effective conductance~$g_m$
\begin{equation}\label{eq:Kirkpatrick54}
\int f(g_0) \frac{g_m - g_0}{g_0 + (z/2 - 1) g_m} \, \mathrm{d}g_0 = 0,
\end{equation}
where $z$ is the valence of nodes in the lattice ($z=4$ for $d=2$, $z=6$ for $d=3$, etc.)

\begin{figure}[!htb]
  \centering
  \includegraphics[width=\columnwidth]{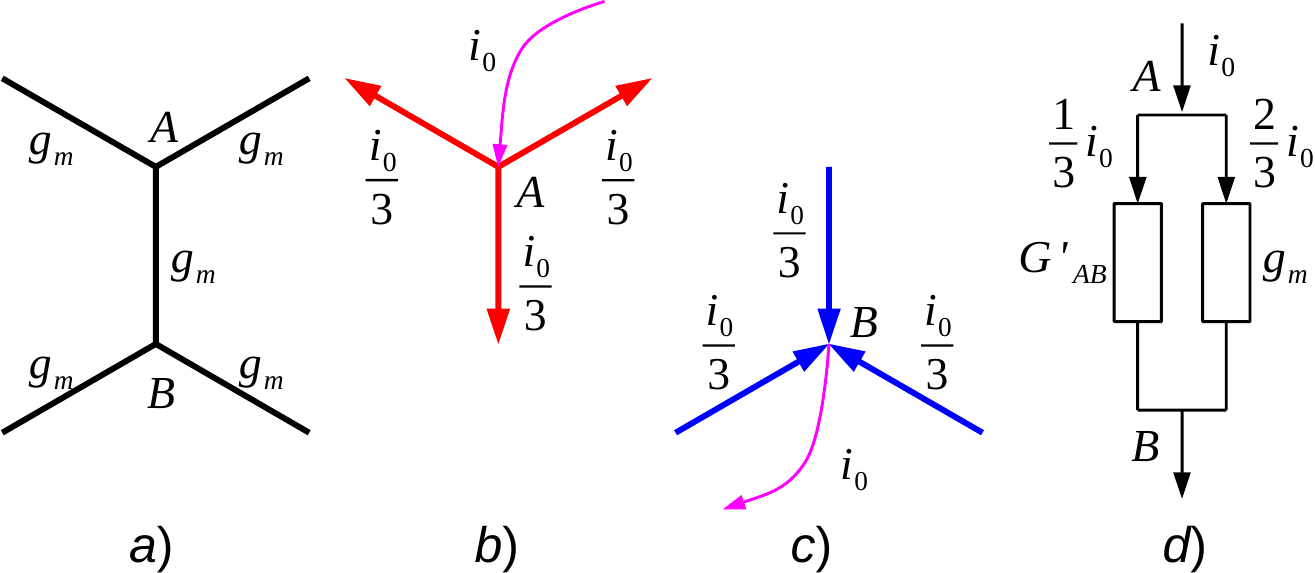}%
  \caption{Schematics to calculation of the conductance between two nearest nodes of a uniform hexagonal network.\label{fig:network6conductivity}}
\end{figure}

Following~\cite{Joy1979}, we start with an infinite hexagonal network (honeycomb), in which conductance of each edge is~$g_m$. The conductance $G_{AB}$ between the two nearest nodes $A$ and $B$ of this network (Fig.~\ref{fig:network6conductivity}a) can be found as follows. Let current $i_0$ be injected into node $A$. Due to the symmetry of the system, the current in each of the three edges incident on node $A$ will be equal to $i_0/3$ (Fig.~\ref{fig:network6conductivity}b).  Let current $i_0$ be removed from node $B$. Due to the symmetry of the system, the current in each of the three edges incident on node $B$ will be equal to $i_0/3$ (Fig.~\ref{fig:network6conductivity}c). Then, by virtue of the superposition principle, the total current in the edge $AB$ is equal to $2 i_0 / 3$.

The conductance of the network between nodes $A$ and $B$ is equal to the sum of the conductance of edge $AB$ ($g_m$) and the conductance of the entire network when edge $AB$ is removed ($G'_{AB}$) (Fig.~\ref{fig:network6conductivity}d)
\begin{equation}\label{eq:GAB1}
G_{AB} = g_m + G'_{AB}.
\end{equation}
Since the current through conductance $g_m$ is twice as large as the current through conductance $G'_{AB}$, then $G'_{AB} = g_m/2$. This value is consistent with the general formula for random regular networks~\cite{Marchant1979}.
Hence,
\begin{equation}\label{eq:GAB2}
G_{AB} = \frac{3}{2} g_m.
\end{equation}
In the case of a uniform network, for any node $i$ the following equality holds:
\begin{equation}\label{eq:sumV}
\sum_{j} g_{ij} (V_i - V_j) = 0.
\end{equation}
Here, the summation is carried out over the nodes $j$ adjacent to node $i$. Although the derivation of the two-point resistance was based on the homogeneity of the system, any continuous deformation of the system under consideration cannot affect the electrical conductivity, i.e., the result has to be valid for any system which topologically equivalent to the honeycomb network, in other words, for any planar 3-regular graph (network). This statement is consistent with the results by \citet{Marchant1979}.    

\begin{figure}[!htb]
  \centering
  \includegraphics[width=\columnwidth]{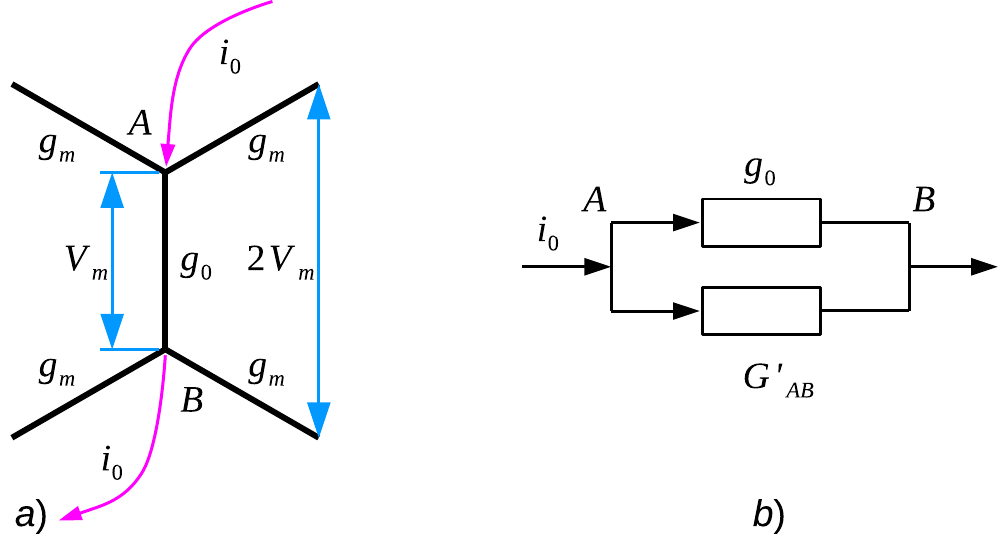}%
  \caption{Sketch to assist calculating the resistance of a hexagonal network.\label{fig:network6conductivity1}}
\end{figure}

Now we replace one of the conductances $g_{AB} =g_m$ with the conductance $g_{AB} = g_0$ (Fig.~\ref{fig:network6conductivity1}a). 
When one conductance is replaced, the solution can be constructed according to the principle of superposition, i.e., to the solution for a uniform field in which the voltages increase by a constant value $V_m$ in each line, we have to add the influence of a fictitious current $i_0$ flowing into node $A$ and flowing out from node $B$,
\begin{equation}\label{eq:i0inA}
2 g_m \frac{V_m}{2} - g_0 V_m = i_0.
\end{equation}
The first term of~\eqref{eq:i0inA} corresponds to the two currents flowing through two conductances $g_m$ into node $A$, while the second one corresponds to the current flowing out of this node via conductance $g_0$. Then, 
\begin{equation}\label{eq:hexinhomi0}
 (g_m - g_0) V_m = i_0.
\end{equation}
This current creates an additional voltage $V_0$ between nodes $A$ and $B$, which can be found using Fig.~\ref{fig:network6conductivity1}b. 
\begin{equation}\label{eq:i0vsV0}
i_0 = (g_0 + G'_{AB}) V_0.
\end{equation}
Taking into account the previously found value for $G'_{AB}$
\begin{equation}\label{eq:V0AB1}
V_0 = \frac{2i_0}{g_m + 2 g_0}.
\end{equation}
Accounting for~\eqref{eq:hexinhomi0}, we have 
\begin{equation}\label{eq:V0AB2}
V_0 = 2 V_m\frac{g_m - g_0 }{ g_m + 2 g_0}.
\end{equation}

Let the conductance values of the edges in a random network obey to the PDF $f_G(g_0)$, then the conductance of the effective uniform network $g_m$ can be found from the requirement that the mean value of $V_0$ has to be equal to zero
\begin{equation}\label{eq:gmhex}
\int_0^{g_\text{max}} f_G(g_0) \frac{g_m - g_0 }{g_m + 2 g_0} \, \mathrm{d}g_0 = 0.
\end{equation}
Equation~\eqref{eq:gmhex} can be solved numerically. For this purpose it can be rewritten in a form convenient for iterations (see Appendix~\ref{sec:appendixiter})
\begin{equation}\label{eq:gmhexiter}
g_m^{(k+1)} = \left[ 3 \int_0^{g_\text{max}} \frac{f_G(g_0)}{g^{(k)}_m + 2 g_0} \, \mathrm{d}g_0 \right]^{-1}.
\end{equation}

\subsection{Numerical results}
Numerical solution of \eqref{eq:gmhexiter} leads to the values of the effective electrical conductance presented in Table~\ref{tab:gm}. To check the accuracy, the average value of potential fluctuation, $\langle V_0 \rangle$, is presented. 
$a = \sqrt{2 (3 n_s\sqrt{3})^{-1} }$
is the cell edge length of a regular hexagonal network with cell density $n_s$; $g_1 = g_m/a$ is the conductance per unit length, $\sigma_\text{hex}= g_m/\sqrt{3}$ is the electrical conductance of a honeycomb network with edge conductance~$g_m$. 
\begin{table}[!htb]
  \centering
  \caption{Effective electrical conductance for different seed densities, $n_s$. }\label{tab:gm}
\begin{ruledtabular}
  \begin{tabular}{ccccccc}
    $n_s$ & $\sqrt{n_E}$ & $g_m$ & $\langle V_0 \rangle \times 10^5$ & $a$ & $g_1$ & $\sigma_\text{hex}$\\
    \hline
0.01 & 0.173 & 0.176 & 1.85 & 6.204 & 0.0283 & 0.1014 \\
0.25 & 0.866 & 0.878 & 1.86 & 1.241 & 0.7075 & 0.5068 \\
0.64 & 1.386 & 1.405 & 1.85 & 0.776 & 1.8111 & 0.8109 \\
1.0  & 1.732 & 1.756 & 1.85 & 0.620 & 2.8298 & 1.0136 \\
2.0  & 2.449 & 2.483 & 1.85 & 0.439 & 5.6596 & 1.4335 \\
3.0  & 3.000 & 3.041 & 1.85 & 0.358 & 8.4894 & 1.7556 \\
4.0  & 3.464 & 3.511 & 1.85 & 0.310 & 11.319 & 2.0272 \\
5.00 & 3.873 & 3.926 & 1.85 & 0.277 & 14.149 & 2.2665 \\
  \end{tabular}
  \end{ruledtabular}
  \end{table}

\begin{figure}[!htb]
  \centering
  \includegraphics[width=\columnwidth]{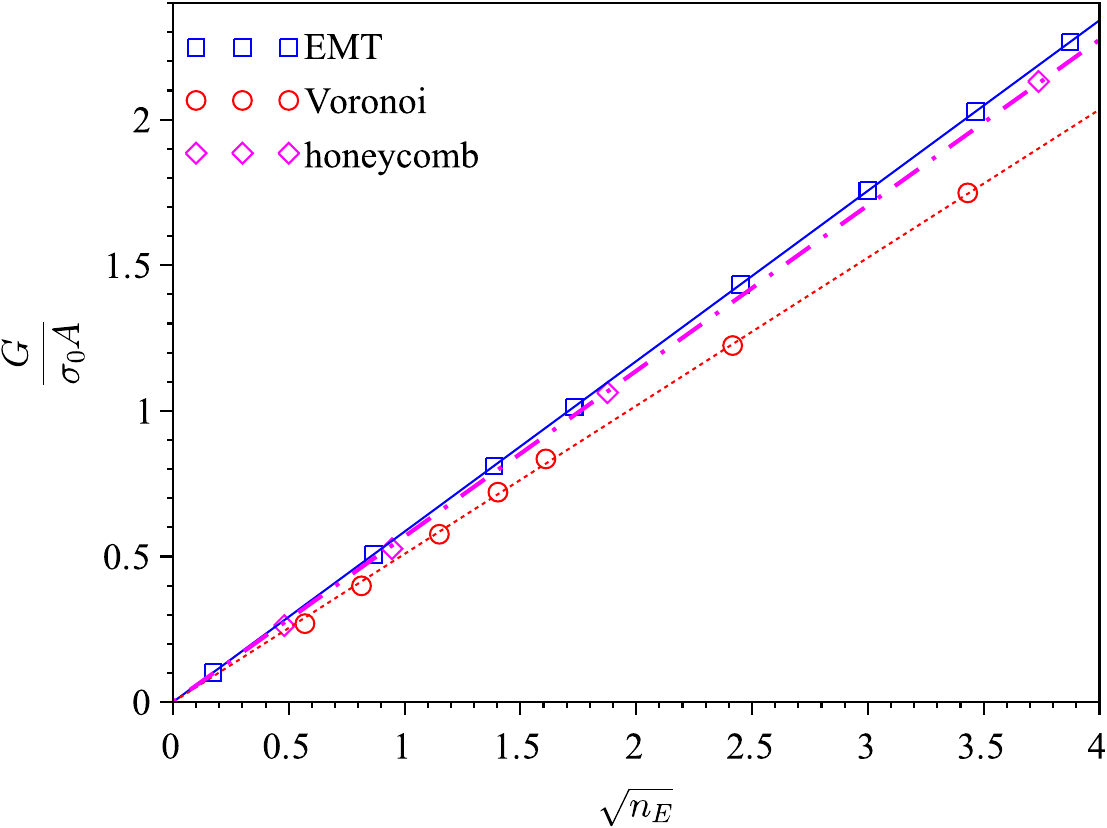}
  \caption{Comparison of the results of direct computations of the electrical conductance of random hexagonal networks as well of random Voronoi networks~\cite{Tarasevich2023a} with predictions of the EMT. Lines correspond to least squares fits.\label{fig:simvsEMT}}
\end{figure}

Figure~\ref{fig:simvsEMT} shows a comparison of the results of direct computations of the electrical conductance of the random Voronoi network with the predictions of the EMT. The EMT predicts a linear dependence of the electrical conductance on the square root of the edge concentration $\sigma = 0.5852 \sqrt{n_E}$ (solid line in Fig.~\ref{fig:simvsEMT}), but slightly overestimates the electrical conductance as compared to direct calculations, which give a slope of $0.5087 \pm 0.0027$~\cite{Tarasevich2023a} (dotted line), i.e., the EMT predicts the value of slope about 15\% larger as compared to direct computations. The EMT is slightly worse as compared to the MFA (slope $\approx 0.57735$)~\cite{Tarasevich2023a}. 

It should be kept in mind that the Voronoi diagram is not a perfect hexagonal network, viz, its cells have 6 nodes only in average.  To estimate the effect of network imperfection, we computed the electrical conductances of hexagonal networks in which the resistivities of edges correspond to the PDF of edge length in random Voronoi diagrams~\cite{Brakke2005}. In this case the slope is 0.5686 (dash-dot line). This small deviation of the EMT prediction and the direct computation is not surprising, since~\citet{Marchant1979} have explained  the nature of the error inherent in the application of EMT to regular networks. Namely, replacement of the distribution of edge conductivities by a unique value found in the `effective network' introduces an error that grows with the broadening of the real distribution of conductivities; i.e., the broader the PDF the larger the error.

\section{Conclusion\label{sec:concl}}

Following~\cite{Zeng2020,Kim2022,Tarasevich2023,Esteki2023,Tarasevich2023a}, we mimic crack-template-based transparent conductive films, using the random resistor networks. In our study, this random resistor network corresponded to a hexagonal network with random conductivities of edges. To estimate the dependence of the electrical conductance of such transparent conductive films on main physical parameters, we utilised the effective medium theory in the same manner as~\cite{Kirkpatrick1971,Kirkpatrick1973,Joy1978,Joy1979}. For random hexagonal networks, the effective medium theory provides a nice approximation of the dependency of the electrical conductivity on the number density of edges. We found that the electrical conductance is $0.5852 \sqrt{n_E}$. Comparison of the effective medium theory predictions with direct computations of the dependency of the electrical conductance on the number density of edges in the random plane Voronoi tessellation, suggests that not only valence of vertices preordains the main behavior of the network conductance. However, even an advanced consideration~\cite{Marchant1979} based on Foster's theorem~\cite{Foster1961} can hardly improve the prediction significantly.

\appendix*
\section{Transformation of the equation for the electrical conductance of the effective medium to a form suitable for iteration\label{sec:appendixiter}}
Consider~\eqref{eq:gmhex}. Since $g_m - g_0$ can be presented as $\frac{1}{2}\left[3g_m - (g_m + 2g_0) \right]$,
then
\begin{multline}\label{eq:appAtransf}
 \frac{g_m - g_0 }{g_m + 2 g_0} = \frac{1}{2}\frac{3g_m - (g_m + 2g_0) }{g_m + 2 g_0}\\  = \frac{1}{2}\left[\frac{3g_m}{g_m + 2 g_0} - 1\right].
\end{multline}
Hence, \eqref{eq:gmhex} can be rewritten as follows
\begin{multline}\label{eq:appAmeanV0}
  \int_0^{g_\text{max}} f_G(g_0) \frac{g_m - g_0 }{g_m + 2 g_0} \, \mathrm{d}g_0 \\
  = \frac{1}{2} \int_0^{g_\text{max}} f_G(g_0) \left[3\frac{g_m }{g_m + 2 g_0} - 1 \right] \, \mathrm{d}g_0 = 0.
\end{multline}
According to definition of a PDF, $\int_0^{g_\text{max}} f_G(g_0) \, \mathrm{d}g_0 = 1,$
hence, 
\begin{equation}\label{eq:apppAgmtrans}
   3g_m \int_0^{g_\text{max}} \frac{ f_G(g_0)}{g_m + 2 g_0} \, \mathrm{d}g_0 = 1 .
\end{equation}
  Equation
\begin{equation}\label{eq:apppAgm}
 g_m = \left(3\int_0^{g_\text{max}} \frac{ f_G(g_0)}{g_m + 2 g_0} \, \mathrm{d}g_0 \right)^{-1}
\end{equation}
  is suitable to find $g_m$ iteratively
\begin{equation}\label{eq:apppAgkiter}
 g^{(k+1)}_m = \left(3\int_0^{g_\text{max}} \frac{ f_G(g_0)}{g^{(k)}_m + 2 g_0} \, \mathrm{d}g_0 \right)^{-1}.
\end{equation}

Then, the formula to calculate the effective conductance can be rewritten using $f_G$ as follows
\begin{multline}\label{eq:appint}
  \int_0^{g_\text{max}} \frac{ f_G(g_0)}{g^{(k)}_m + 2 g_0} \, \mathrm{d}g_0 =
  g_1 \int_0^{g_\text{max}} \frac{ f_l\left( \frac{g_1}{g_0} \right)}{g^{(k)}_m  + 2 g_0} \, \frac{\mathrm{d}g_0}{g_0^2}
\\
  =\int_0^{l_\text{max}} \frac{ f_l\left( l \right)}{g^{(k)}_m + 2 g_1 l^{-1} } \, \mathrm{d}l,
  \end{multline}
  hence,
\begin{equation}\label{eq:appgk11}
g^{(k+1)}_m = \left(3 \int_0^{l_\text{max}} \frac{ f_l\left( l \right)}{g^{(k)}_m + 2 g_1  l^{-1}} \mathrm{d}l\right)^{-1}.
\end{equation}
Assuming $g_1 = 1$, we have
\begin{equation}\label{eq:appgk1}
g^{(k+1)}_m = \left(3 \int_0^{l_\text{max}} \frac{ f_l\left( l \right)}{g^{(k)}_m + 2 l^{-1}} \mathrm{d}l \right)^{-1}.
\end{equation}
  Accordingly, the average fluctuation of the potential can be rewritten as follows
\begin{multline}\label{eq:meanV0vslhex}
\langle V_0 \rangle =  \int_0^{g_\text{max}} \frac{g_1 }{g_0^2} f_l\left( \frac{g_1}{g_0} \right) \frac{g_m - g_0 }{g_m + 2 g_0} \,  \mathrm{d}g_0\\ 
=\int_{0}^{l_\text{max}} f_l( l ) \frac{g_m - g_1 l^{-1} }{g_m + 2 g_1 l^{-1}} \, \mathrm{d}l = 0.
\end{multline}
Since $g_1 = 1$, then 
\begin{equation}\label{eq:meanV0vslhex1}
\langle V_0 \rangle = \int_{0}^{l_\text{max}} f_l( l ) \frac{g_m l - 1}{g_m l + 2 } \, \mathrm{d}l = 0.
\end{equation}


\begin{acknowledgments}
We acknowledge funding from the FAPERJ, Grants No.~E-26/202.666/2023 and No.~E-26/210.303/2023 (Y.Y.T. and F.D.A.A.R) and from the Russian Science Foundation, Grant No. 23-21-00074 (I.V.V. and A.V.E.). Y.Y.T. thanks Prof. Ken Brakke for explanation of some points in~\cite{Brakke2005}.
\end{acknowledgments}

\bibliography{EMTarticle}

\end{document}